\documentclass[letterpaper, 10 pt, conference]{ieeeconf}  % Comment this line out
\IEEEoverridecommandlockouts                              % This command is only
\usepackage{amsmath}
\usepackage{graphicx}
\usepackage{amssymb}
\usepackage{epstopdf}
\usepackage{url}
\usepackage{mathabx}
\usepackage{color}

\author{Michael Zargham$^{1}$, Zixuan Zhang$^{2}$, Victor Preciado$^{2}$% <-this % stops a space
\thanks{$^{1}$Michael Zargham is Founder \& CEO at BlockScience. Email:
        {\tt\small zargham at block.science}}%
\thanks{$^{2}$Zixuan Zhang and Victor Preciado are with the Department of Electrical and Systems
Engineering, University of Pennsylvania. Email:
        {\tt\small $\{\hbox{zixuanzh, preciado}\}$  at seas.upenn.edu}}%
}%

\usepackage[ruled]{algorithm2e} % For algorithms

\SetAlFnt{\small}
\SetAlCapFnt{\small}
\SetAlCapNameFnt{\small}
\SetAlCapHSkip{0pt}
\IncMargin{-\parindent}

\newtheorem{definition}{Definition}

\newtheorem{conjecture}{Conjecture}
\newtheorem{result}{Result}

\title{A State-Space Modeling Framework for Engineering Blockchain-Enabled Economic Systems}

\begin{document}
\maketitle

% Title portion
\begin{abstract}

%\textcolor{red}{MIKE, please take a shot at the Abstract and I will revise afterwards. Try to keep it to around half a column...}

Decentralized Ledger Technology (DLT), popularized by the Bitcoin network, aims to keep track of a ledger of valid transactions between agents of a virtual economy without the need of a central institution for coordination. In order to keep track of a faithful and accurate list of transactions, the ledger is broadcast and replicated across the machines in a peer-to-peer network. To enforce that the transactions in the ledger are valid (i.e., there is no negative balance or double spending), the network ‘as a whole’ coordinates to accept or reject new transactions according to a set of rules aiming to detect and block the operation of malicious agents (i.e., Byzantine attacks).
Consensus protocols are particularly important to coordinate the operation of the network, since they are used to reconcile potentially conflicting versions of the ledger. Regardless of the architecture and consensus mechanism used, the resulting economic networks remains largely similar, with economic agents driven by incentives under a set of rules.
Due to the intense activity in this area, proper mathematical frameworks to model and analyze the behavior of blockchain-enabled systems are essential. In this paper, we address this need and provide the following contributions: (i) we establish a formal framework, using tools from dynamical systems theory, to mathematically describe the core concepts in blockchain-enabled networks, (ii) we apply this framework to the Bitcoin network and recover the key properties of the Bitcoin economic network, and (iii) we connect our modeling framework with powerful tools from control engineering, such as Lyapunov-like functions, to properly engineer economic systems with provable properties.
Apart from the aforementioned contributions, the mathematical framework herein proposed lays a foundation for engineering more general economic systems built on emerging Turing complete networks, such as the Ethereum network, through which complex alternative economic models are being explored.

\end{abstract}

\section{Introduction}

%CONTEXT PARAGRAPH ACCESSIBLE TO THE NON-EXPERT
During the 2007-2008 global financial crisis, serious abuses by major financial institutions initiated a series of events resulting in a collapse of the loosely regulated financial network. This crash unveiled major weaknesses of traditional financial system and instigated a general feeling of distrust on banking institutions. In this context, decentralized economic systems based on cryptocurrencies, such as Bitcoin \cite{nakamoto2008bitcoin}, were developed and launched by a group of cryptographic activists who believed in social change through censorship-resistant and privacy-enhancing technologies. Prior to Bitcoin, several attempts to establish digital currencies were made, including  \emph{b-money} by Dai \cite{dai1998bmoney}, \emph{hashcash} by Finney \cite{back2002hashcash}, and \emph{bit gold} by Szabo \cite{szabo2005bitgold}. Instead of relying on large financial institutions for their operation, these cryptocurrencies proposed a decentralized economy in which a collection of economic agents coordinate through a peer-to-peer networks of computers via a blockchain protocol.

At a high level, blockchain aims to keep track of a ledger of valid transactions between agents of the economy without the need of a central institution for coordination. In order to keep track of a faithful and accurate list of transactions, the ledger is broadcast and replicated across all the machines in a peer-to-peer network. To enforce that the transactions in the ledger are valid (i.e., there is no negative balance or double spending), the network `as a whole' coordinates to accept or reject new transactions according to a set of rules aiming to detect and block the operation of malicious agents (i.e., Byzantine attacks). Blockchain implements this idea by bunching together a group of new transactions that are added into a chain of blocks only if these transactions are validated by the peer-to-peer network. Consensus protocols are particularly important in this validation step, since they are commonly used to reconcile conflicting versions of the ledger. A particular protocol used to enforce consensus is \emph{Proof of Work (PoW)} \cite{bentov2014cryptocurrencies}, currently used in Bitcoin and Ethereum. PoW is just one particular example of many other consensus protocols, such as the Practical Byzantine Fault Tolerant algorithm (PBFT) \cite{castro1999practical}, Proof of Stake (PoS) \cite{ethereum-pos}, or Delegated Proof of Stake (DPoS) \cite{bitshares-dpos}, to mention a few. These consensus protocols differ by their degree of decentralization, fault tolerance, throughput, and scalability \cite{consensus-review}. 

It is worth mentioning that, beyond applications in cryptocurrency, blockchain can be used to store other forms of data on the network \cite{bartoletti2017analysis}. This capability has inspired a plethora of protocols and applications aiming to, for example, certify the existence \cite{factom} and tracking of asset ownership \cite{omni-layer}. A notable application is Ethereum, a blockchain with a Turing-complete programming language \cite{buterin2014ethereum}. Ever since the inception of Ethereum, there has been an explosion of tokens sitting on top of the Ethereum network \cite{bat-token, zrx-token, gridplus-token}. Regardless of the architecture and consensus mechanism used, the resulting economic networks remains largely similar, with economic agents driven by incentives under a set of rules. Recent efforts have focused on implementing these currencies and other business logic such as decentralized exchanges through on-chain programs called smart contracts, first defined by Szabo in the 1990's \cite{szabo1994smart,szabo1996smart}. 

%WHAT WE DO
Due to the intense recent activity in this area, proper mathematical frameworks to model and analyze the behavior of blockchain-enabled systems are essential. In this direction, we find notable work by DeFigueiredo and Barr on path independence \cite{1524055} and Dandekar et al. on credit networks \cite{Dandekar:2011:LCN:1993574.1993597}. These works lay down the theoretical foundation for projects like Ripple \cite{schwartz2014ripple} and Stellar \cite{mazieres2015modeling}. The main objective of this paper is to propose a new mathematical framework, based on tools from dynamical systems theory and control engineering, to model and analyze the function of blockchain-enabled systems. Before we provide more details about our particular modeling framework, let us describe the blockchain from a network point of view. In particular, the blockchain is comprised of two different networks: (a) a peer-to-peer computer network whose objective is to enforce and broadcast a valid state of the ledger, and (b) a network of valid transactions whose edges represent transactions and whose nodes are unique addresses used to encode the source and destination of a particular transaction. The main objective of the blockchain protocol is to keep a valid, up-to-date copy of the structure of network b) in all the nodes in network a).

As we mention above, we use tools from dynamical systems theory to model the function of the blockchain which is independent of the underlying computational infrastructure implemented. In particular, we borrow a modeling framework widely used in control engineering called \emph{state space representation} \cite{sontag2013mathematical}. According to this framework, there is a set of abstract variables, called states, evolving over time (either continuous or discrete) according to a set of rules. In the discrete-time case, the evolution rules are described in terms of a first-order difference equation, in which the values of the states at a given time $t\in \mathbb{N}$ depend exclusively on the values of the states at time $t-1$. Hence, given the initial values for the states at the origin of time (i.e., $t=0$), it is possible to recover the states at any time $t>0$ by solving this recursion. There is a rich mathematical theory to analyze state-space models, specially in the so-called linear case, in which the states at time $t$ depend on the states at time $t-1$ according to a linear transformation. In this paper, we propose a linear state-space model of the blockchain network whose set of states represent transaction addresses. Notice that, as new transactions take place in the decentralized economy, the number of states in the model increases monotonically over time. This results in some technical difficulties that we overcome by proposing a linear time-expanding (LTE) state-space model which we will use to analyze the temporal behavior of the blockchain. As a reference case, we will model and analyze the evolution of the state in the public Bitcoin network using the aforementioned LTE modeling framework. Using tools from state-space theory (in particular, Lyapunov-like functions \cite{khalil1996noninear}), we will illustrate how to enforce a global property, namely, the total amount of currency in the system, using local state-transition rules.

The paper is structured as follows: After providing appropriate background, we describe the LTE state-space model in Section \ref{sec:Notation}. In Section \ref{sec:Bitcoin}, we analyze the behavior of the Bitcoin network using our framework and provide tools for optimization and control enabled by our framework in Section\ref{sec:Control}. We finalize with some conclusions and future considerations in Section \ref{sec:Conclusions}.

\section{Theoretical Framework} \label{sec:Notation}
%The following characterization is a model in so far as it is not an attempt to describe how a blockchain network works but rather creates a useful abstraction of its properties to engineer economic systems within such a network. As part of this abstraction all state variables are treated as real valued fields. It is the authors belief that relaxing this assumption is possible but that the derivations under this assumption are more intuitive and straightforward. 
The following characterization attempts to create a useful abstraction over the properties of a blockchain network. It is not an attempt to describe how a blockchain network works, but to provide tools to engineer economic systems within such a network. All state variables are real-valued to make derivations more intuitive and straightforward.

\subsection{Characterizing the Ledger}

% \begin{definition} The \textbf{Ledger State} is the shared data structure $\mathbf{B}(k)\in\mathcal{B}$ of the blockchain Network which evolves in discrete time denoted by $k$ and $\mathcal{B}$ denotes the domain of the ledger, which space of all valid ledger states $\mathbf{B}(k)$ can take for any $k$.
% \end{definition}

\begin{definition} The \textbf{Ledger State} is a shared data structure $\mathbf{B}(k)\in\mathcal{B}$ of a blockchain network which evolves in discrete time denoted by $k\in \mathbb{N}$. $\mathcal{B}$ denotes the domain space of all valid ledger states $\mathbf{B}(k)$ for any $k$.
\end{definition}

% The Ledger State evolves in discrete time $k$, according to the rules of accounts which are under the control of the possessors of the private keys to those accounts. The set of all accounts at block $k$ is $\mathcal{A}_k$, and the ledger state can be partitioned as $\mathbf{B}(k) = \bigtimes_{a\in \mathcal{A}_k } \mathbf{B}_a(k)$ where $\mathbf{B}_a(k)$ is the state of account $a$; the $\bigtimes$ symbol denotes the generalized cartesian product. 

The Ledger State evolves in discrete time $k$, according to the rules and actions of all accounts. Each account is only accessible through the ownership of its private key. The Ledger State can thus be partitioned as $\mathbf{B}(k) = \bigtimes_{a\in \mathcal{A}_k } \mathbf{B}_a(k)$ where $\mathcal{A}_k$ denotes the set of all accounts at block $k$ and $\bigtimes$ denotes the generalized cartesian product.

%note this definition should subsume both internal and external accounts in the ethereum sense. It can be thought of as an on-chain stateful micro-service with a public key and a private key for proving ownership which would be required to use methods only accessible to the owner. This is tricky -- making a first pass but will definitely need attention and rework.
% \begin{definition}\label{account}
% An \textbf{Account} is a unique element of the ledger, identified by an address $pk(a)$ which has an associated $sk$ required for proof of right to modify the code defining the account. 
% \end{definition}

\begin{definition}\label{account}
An \textbf{Account} is a unique element in a ledger, identified by a public address $pk(a)$ with an associated private key $sk$. $sk$ is required as a proof of right to both modify code defining the account and perform actions defined by the code.
\end{definition}

% An account may contain code defining its state variables, external methods for interacting with other accounts, and other supporting methods which are used internally. Since public keys are the unique identifiers of accounts, linking to the conventional notation of the cryptography community $PK=\mathcal{A}_k$ is the set of all public keys. 

An account may contain code defining its state variables, external methods for interacting with other accounts, and other internally-used supporting methods. Since public keys are the unique identifiers of accounts, $PK=\mathcal{A}_k$ is the set of all public keys, similar to notation used in the cryptography community. 

\begin{definition}
A \textbf{Method} is a state transition function
\begin{equation}
f:(\mathcal{U}, \mathcal{X}) \rightarrow \mathcal{X}
\end{equation} 
where $\mathcal{U}=\mathcal{U}(x)$ for $x\in \mathcal{X}$ is the set of legal actions.
\end{definition}

% The set of methods provided by account $i$ that are available to account $j$ is denoted $\mathcal{F}_{(i,j)}$ with elements of the form 
% \begin{equation}
% f_l:(\mathcal{U}_l, \mathcal{X}) \rightarrow \mathcal{X}
% \end{equation} 
% where $l$ denotes the integer index of the functions within the set. \textcolor{red}{Commonly, the actions $\mathcal{U}_l(x)$ available to account $j$ will depend explicitly on $x_i$.\\}

Each method is defined by a particular account and is made available to a set of accounts. The set of methods provided by account $i$ that are available to account $j$ is denoted $\mathcal{F}_{(i,j)}$ with elements of the form 
\begin{equation}
f_l:(\mathcal{U}_l, \mathcal{X}) \rightarrow \mathcal{X}
\end{equation} 
where $l$ denotes the integer index of the functions within the set. The actions $\mathcal{U}_l(x)$ available to account $j$ may depend explicitly on $x_i$.

A holder of the private key for account $a$ can take any action by using any method in the set 
\begin{equation}
   \mathcal{F}_a = \bigcup_{i}\mathcal{F}_{(i,a)} 
\end{equation}
which has the associates space of actions
\begin{equation}
    \mathcal{U}_a(x) = \bigcup_{l\in \mathcal{F}_a} \mathcal{U}_l(x)
\end{equation}
for any $x\in\mathcal{X}$. %A private key holder of account $a$, or another account may take actions $u\in \mathcal{U}_a$

%\textcolor{red}{
%Notes -- need to refactor the way these methods are defined so that we are not confusing the concepts\\
%-- any method has an account that provides that method\\
%-- and set of accounts to which that method is exposed. \\
%}
%Any particular account $a$ has internal state $x_a(k) \in \mathcal{X}_a$ for any block $k$ where $\mathcal{X}_a$ is the set of all legal actions as determined by the account's codebase. A private key holder of account $a$, or another account may take actions $u\in \mathcal{U}_a$ on account $a$ where $\mathcal{U}_a$ is the actions which are well defined for account $a$. These actions are associated with methods $\mathcal{F}_a$ with elements of the form 
%\begin{equation}
%f_l:(\mathcal{U}_l, \mathcal{X}) \rightarrow %\end{equation} 
%where $l$ denotes the integer index of the function within the set and $x\in \mathcal{X}$ is state of all accounts, and  $\mathcal{X} =\bigtimes_{a\in \mathcal{A}_k} \mathcal{X}_a$ is the space of all valid states of those accounts. The methods available to the private key holder of account $a$ may differ from those available to other accounts.

% For the purpose of discussion, it is not required to further define internal operations not related to interactions with other accounts due to the explicit assumption that the accounts are passive, implying that any state changes which occur internally can be adequately accounted for in the mappings in $\mathcal{F}_a$.

For the purpose of discussion, it is not required to further define internal operations that do not involve other accounts. Accounts are assumed to be passive explicitly and hence any internal state changes can be adequately accounted for in mappings $\mathcal{F}_a$. 

The Ledger State directly related to account $a$ at block $k$ is 
\begin{equation}
   \mathbf{B}_a(k) = \{x_a(k), \mathcal{T}_a(k)\} 
\end{equation}
where $\mathcal{T}_a(k)$ is an ordered list of all transactions, involving account $a$ in block $k$. Ordering is required as earlier changes to the state may  influence the validity of later transactions in the sequence.

%for now characterizes a single signer transactions, there will be other work to resolve the definitions required for multi-sig transactions and/or multi-sig accounts
%two approaches -- 1 multi-sig versions are generalizations of these or alternatively they could be constructed as compositions
\begin{definition} A \textbf{Transaction} denoted by $\mathbf{tx}$ is the list of state changes caused by a discrete action $u\in \mathcal{U}_a$ for some account $a\in A(k)$. The initiating account is denoted $a_0$ and the other accounts whose states are impacted by the transaction are denoted by $a_i$ for $i\in{1,2,3,...n}$; therefore the transaction itself is defined as the list $\mathbf{tx} = [\Delta x_{a_0},\Delta x_{a_1},\ldots ,\Delta x_{a_n} ]$ where $\Delta x_{a_i}$ is the change in state of account $a_i$ as a result of the action $u$ represented by the transaction $\mathbf{tx}$. 
\end{definition}

Under this notation, state transitions within a transaction are atomic and intermediate state transitions that may occur as part of a smart contract are not accounted for. A finer grained model would be required to handle that level of precision.

The simplest transaction occurs when $u\in \mathcal{U}_{a_0}$. The transaction only modifies the state of the initiating account by legally calling a method $f\in\mathcal{F}_{a_0}$ that does not change the state of any other accounts. The transaction is completely characterized by $\Delta x_{a_0}$ and can be validated by evaluating $f(u, x)$ within the sequence of transactions $\mathcal{T}_{a_0}(k)$ and showing that $x_{a_0}+\Delta x_{a_0} = f(u, x)|_{a_0} \in \mathcal{X}_{a_0}$. The notation $f(\cdot)|_{a}$ denotes the element of the output of $f(\cdot)$ associated with account $a$.

A more interesting case is when account $a_0$ takes an action $u\in \mathcal{U}_{a_o}$ which changes the state of other accounts $a_i\in \mathcal{A}_k$. In order for $\mathbf{tx}$ to be a valid transaction the resulting state transitions for each other account must be valid. If the function $f$ is implemented by calling other methods, the validity only holds if all methods are properly applied.

%$x_{a_i}+\Delta x_{a_i} \in \mathcal{X}_{a_i}$ for all impacted accounts $a_i$ which requires checking the sequence $\mathcal{T}_{a_i}(k)$ for each $a_i$.
%The final basic case is when the account $a_0$ takes an action $u\in\mathcal{U}_{a}$ for some $a\neq a_0$. That is to say the method $f_a(u, x)$ is called by account $a_0$ but is provided by $a$. It is assumed that any function $f_a(u, x)$ may have restricted access by initiating account, the property to verify is where the account $a_0$ is qualified to take action $u$ associated with function $f_a(u, x)$ given accounts states $x$. Once it is determined that $a_0$ is qualified to take the action, the characterization reverts to the prior case; 
The rules on domain $\mathcal{X}$ are directly characterized by the range of state transition functions provided by accounts so it suffices to check the computation of $f(u, x)$ to assert the validity of each output account state: 
\begin{eqnarray*}
x_{a_0}+\Delta x_{a_0}=f_{a}(u, x)|_{a_0}&\in&\mathcal{X}_{a_0},\\
x_{a_1}+\Delta x_{a_1}=f_{a}(u, x)|_{a_1}&\in&\mathcal{X}_{a_1},\\
\ldots &&\\
x_{a_n}+\Delta x_{a_n}=f_{a}(u, x)|_{a_n}&\in&\mathcal{X}_{a_n}.
\end{eqnarray*}
%irrespective of which account $a_i$ is the function provider account $a$.

\begin{definition} A \textbf{Transaction Block} is an ordered list of transactions $\mathbf{TX}(k) = [\mathbf{tx}_0, \mathbf{tx}_1, \ldots, \mathbf{tx}_m]$. Since transactions are lists of account state changes, it can interpreted as a flat list. A block is valid if each individual transaction is computed correctly and the state change for each account after each sequential transaction is valid. 
\label{transactionblock}
\end{definition} 

Consider a block $k$ with prior account states $x(k-1)$, the global account state update is given by
\begin{equation}
    x(k) = x(k-1) + \Delta x(k)
\end{equation}
where $\Delta x$ is the global account state update achieved by all transactions in $\mathbf{TX}(k)$.
\begin{equation}
    \Delta x_a(k) = x_a(k) - x_a(k-1)=\sum_{i}\Delta x^{(i)}_a, \, \forall a\in \mathcal{A}_k \label{decomp}
\end{equation}
where the index $i$ denotes the indices of all $\mathbf{tx}\in \mathcal{T}_a(k)\subseteq\mathbf{TX}(k)$, applying the definition $\mathcal{T}_a(k)$ is the set of transactions at block $k$ including state changes to account $a$. The global state update
\begin{equation}
    \Delta x(k) = x(k)- x(k-1) 
\end{equation}
can be completely characterized by $\mathbf{TX}(k)$ and evaluated account-wise according to equation \eqref{decomp}.

\begin{definition}
A \textbf{Block} is the ledger state $\mathbf{B}(k)$ at $k$ which given the definitions above can be formally characterized as the pair
\begin{equation}
    \mathbf{B}(k) = (x(k), \mathbf{TX}(k) ).
\end{equation}
\end{definition}

The underlying mechanisms that maintain the state of the blockchain do so by keeping the transaction history. This makes a direct connection to dynamical systems theory
\begin{equation}
x(K) = x(0)+ \sum_{k=1}^{K} \Delta x(k) \label{sums}
\end{equation}
where $x(0)$ denotes the initial state, also referred to as the genesis block.

\begin{result}
The evolution of the Ledger State $\mathbf{B}(k)$ for $k=0,1,2, \ldots$ is the trajectory of a discrete time second order networked system. 
\end{result}

It is a networked system comprised of accounts $a$ with locally defined internal dynamics and rules for interacting according to the account Definition \ref{account}. It is a discrete second order system because the ledger state $\mathbf{B}(k)$ contains precisely the states $x_a(k)$ and the backward discrete derivatives $\Delta x_a(k) = x_a(k) - x_a(k-1)$ for all accounts $a \in \mathcal{A}_k$.

Any mechanism that can be implemented as an account under this framework provides an explicit contribution to the actions available to all other accounts within the system. The explicit characterization of an account and its subsequent state changes permits the estimation of changes in any utilities defined over the network state. Using the second order discrete networked system model, it is possible to both formally analyze the reachable state space and simulate the response to incentives with respect to a variety of behavioral assumptions.

\subsection{Characterizing the Peer-to-Peer Network}

Under this formal model there are two distinct concepts matching the term \textbf{Network}. The state space model defines the evolution of a network of interacting accounts. From this point of view, the economic network is a robotic network with agents represented by accounts, each of which has its own unique state space and action space defined in part by all of the other agents (accounts) in the network. The agent (account) states of all network participants and their backward discrete derivatives are visible to other agents and any external observer capable of querying the Ledger State.

Consideration of the external viewer brings attention to the other concept of a network which is required to model this system; the communication and computation network responsible for maintaining account states, computing state updates, verifying the validity of blocks of transactions, and to agree on the correct sequence of blocks when multiple valid sequences are available.

\begin{definition}
A \textbf{Node} is a member of the Peer-to-Peer Network with the ability to broadcast a transaction $\mathbf{tx}$ for which it can prove control of the initiating account $a_0$ using the associated private key, and the ability to verify the validity of transactions broadcast by other nodes.
\end{definition}

\begin{definition}
The \textbf{Peer-to-Peer Network} is the set of nodes $j\in \mathcal{V}$, participating in the communication and computation network, each maintaining a copy of the Ledger State $\mathbf{B}_j(k)$ and edges in this network represent communication between nodes.  
\end{definition}

Note that each node $j$ may have its belief of the Ledger State $\mathbf{B}_j(k)$ such that for any two nodes $\mathbf{B}_j(k) \neq \mathbf{B}_{j'}(k)$ for $j\neq j'$. However, It is guaranteed by the underlying cryptographic protocol that both $\mathbf{B}_j(k), \mathbf{B}_{j'}(k) \in \mathcal{B}$. 

\begin{definition} 
A \textbf{Chain} is a valid sequence of Ledger States, $\mathbf{C}(K) =[\mathbf{B}(k)\in \mathcal{B}$ for $k=0,1,\ldots,K] \in \mathcal{B}^{K+1} $ where $\mathbf{B}(0)$ is the genesis block and $K$ is the block height.
\end{definition}

The cryptographic protocol maintaining the ledger uses sequences of hashing functions to maintain a strict ordering on blocks such that any attempt to manipulate the history of the ledger state is immediately detectable by all nodes in the communication and computation network. Since the cryptographic protocol only accepts blocks for which all state transitions are defined by legal transactions, the chain is also guaranteed to contain a self consistent historical trajectory of the state space model, both states and derivatives, starting with the initial condition $x(0)$ as defined in the genesis block. 

Returning to the issue of nodes maintaining valid but different chains $\mathbf{C}_j(K_j) \neq \mathbf{C}_{j'}(K_{j'})$ which conclude with Ledge States $\mathbf{B}_j(K_j), \mathbf{B}_{j'}(K_{j'}) \in \mathcal{B}$ where $K_j$ and $K_{j'}$ may be different block heights.

\begin{definition} The \textbf{Consensus Protocol}, $\mathbb{C}$ is the process by which agents resolve inconsistency:
\begin{equation}
\mathbb{C}: (\mathcal{C}, \mathcal{C}) \rightarrow \mathcal{C}
\end{equation}
returning which of the two otherwise valid chains superseding the other.
\end{definition}

The existence of such a function alone does not ensure the network does not fragment into partitions maintaining conflicting states. To further ensure consensus, the consensus protocol requires the following property.

\begin{conjecture}\label{Consensus}
The Consensus protocol must impose a strict ordering on valid chains $\mathbf{C}\in \mathcal{C}$. It is sufficient that there exists a function 

\begin{equation} 
    \Psi: \mathcal{C}\rightarrow \mathbb{R}
\end{equation}
such that for any $\mathbf{C},\mathbf{C}'\in \mathcal{C}$
\begin{equation}
     \mathbf{C}\neq \mathbf{C}' \implies \Psi(\mathbf{C}) \neq \Psi(\mathbf{C}'). \label{Psi}
\end{equation}
Two nodes may resolve their inconsistency by each setting their Chain to 
\begin{equation}
 \mathbf{C}^* = \hbox{arg}\max_{C\in\{\mathbf{C}, \mathbf{C}'\}} \Psi(C).
\end{equation}
\end{conjecture}

The formalism is consistent with the Nakomoto consensus paradigm where the function $\Psi$ is the amount of work done to reach the current state in the competing chains. While technically it is possible for two chains to have exactly the same amount of work and still differ, but such a discrete event is a measure zero outcome in a continuous probability distribution, thus for the Bitcoin network using total work, equation \eqref{Psi} can be expected to hold with Probability 1.

For the purpose of the economic specification and subsequent design and analysis, it suffices to use any consensus protocol for which Conjecture \ref{Consensus} holds with Probability 1. Using proof schemes such as the one described above results in a lack of finality, meaning that there is always the possibility that another chain will supersede the one a node is maintaining. A consensus algorithm with \textbf{finality}, meaning that agreement on $\mathbf{B}(k)$ for block height $K$ would not be reversible by some later observation, would be by definition Markovian. It would not be required to compare full trajectories $\mathbf{C}(K)$ to come to consensus over $\mathbf{B}(K)$. PoS-based consensus methods under development aim to achieve this property.

%It would be worthwhile to extend the generalization of consensus mechanisms over blockchain networks to formally define and prove sufficient conditions for a finality property that does not rely on waiting for a stochastic process to converge. There are several active threads of research into consensus protocols so this document will simply proceed under the assumption that the communication and computation network is the ultimate arbitrator of the state of the economic network regardless of how consensus is implemented. For the purpose of this document, the analysis proceeds without any assumption of the consensus mechanism but concerns itself with any valid chain containing a valid trajectory of the economic states.

%For the purpose of this document, the state space model will be used to characterize the economic system. This model includes formal set-based constraints which are enforced in the lower level system by cryptographic protocols. The impacts of attacks which successfully violate the set-based constraints can be analyzed in the state-space model via sensitivity analysis of arbitrary objective functions defined over the state space with respect to the constraints those attacks would violate.

\section{Bitcoin Reference Case} \label{sec:Bitcoin}

The public nature of the data in the Bitcoin economic network has made it a great candidate for research on financial flows. These models consider graphs of flows between accounts. This analysis will instead focus structurally on how the very simple rules about what constitutes a valid transaction result in well defined global properties.

\subsection{Linear Time-Expanding Model}

% \textcolor{red}{... arguably its LTV because the state space is expanding by the dynamics themselves are time invariant. Elegantly handing the state space expansion while keeping the dynamics as time invariant is something i find novel about this.}

% Shoud we reframe this as LTV even though only the dimensionality is changing?\\

The Bitcoin economic network is defined over block height $k=0,1,2,\ldots$, and there are $n_k=|\mathcal{A}_k|$ accounts at each block $k$ with the additional caveat that $n_{k+1} \ge n_k$. For consistency of notation with dynamical models on networks, accounts will be referenced with indices $i\in\{1,\ldots, n_k\}$. 
\begin{definition}
A \textbf{Linear Time-Expanding (LTE)} system has a state space model in the form of a discrete time varying linear model with the dimension of the state space $x\in \mathbb{R}^{n_k}$ which is monotonically non-decreasing while the state update matrices vary only in $n_k$.
\end{definition}
%While control of accounts may be lost, the accounts themselves persist. 

Consider a canonical form discrete time linear time varying model:
\begin{equation}
x(k+1) = A_k x(k) + B_k u(k) \label{sys}
\end{equation}
where $x(k) \in \mathbb{R}^{n_k}$.
Under this framework $A_k\in \mathbb{R}^{n_{k+1} \times n_k}$, but since there are no internal dynamics
\begin{equation}
A_k = \left[ \begin{array}{c} I_{n_k}\\ 0  \end{array} \right]\label{aid}
\end{equation}
where $ I_{n_k}$ is the identity matrix. The matrix $B_k$ is an all-to-all incidence matrix encoding all possible sends $B_k\in \{0,1,-1\}^{n_{k+1} \times m_k}$ where $m_k = n_{k+1}\cdot(n_k-1) = |\mathcal{E}_k|$ and $u(k)\in \mathbb{R}^{m_k}$. The edge set is given by $\mathcal{E}_k=\mathcal{A}_{k} \times \mathcal{A}_{k+1}$ because flows must original in accounts that exist at time $k$. 
\begin{equation}
\left[B_k\right]_{ie} = \left\{\begin{array}{ll} 1 & \hbox{if $e=(j,i)$ for any $j$ }\\ -1 & \hbox{if $e=(i,j)$ for any $j$ }\\ 0 & \hbox{otherwise}\end{array} \right. \label{singlespend}
\end{equation}
\begin{figure}[h]
\includegraphics[width=4cm]{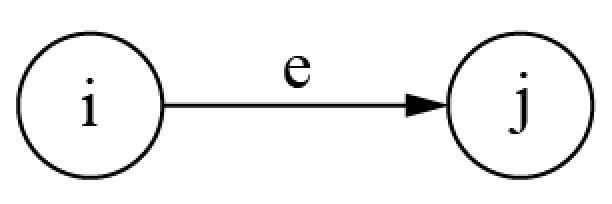}
\centering
\caption{Illustration of incidence matrix. An edge $e$ from $i$ to $j$ represents $i$ sending a transaction to $j$. $B_{i, e} = 1$ and $B_{j, e} = -1$.  }
\centering
\end{figure}
\begin{result}
The system in equation \eqref{sys} is an \textbf{Linear Time-Expanding} (LTE) system because for all $k$, $A_k$ is an augmented identity matrix as defined in equation \eqref{aid} and $B_k$ is an all-to-all incidence matrix as defined in equation \eqref{singlespend}.
\end{result}
The incidence matrix construction enforces the requirement of no double spends. Under this construction the local action $u_e(k) \in \mathcal{U}_e(k)$
\begin{equation}
\mathcal{U}_e(k) = \left\{u_{e=(i,j)} \in\mathbb{R} \big | \sum_j u_{(i,j)} \le x_i(k) \forall i\right\} \label{positivebalances}
\end{equation}
where equation \eqref{positivebalances} enforces the requirement that accounts cannot spend funds they do not have. Note that the requirement is locally enforceable, requiring only the balance of account $i$ and the other transactions to or from account $i$ during block $k$. Viewed from the perspective of account $i$ the local constraint a flow balance
\begin{equation}
    x_i(k) + \sum_{j} u_{(j,i)}(k)-u_{(j,i)}(k) \ge 0. \label{netflows}
\end{equation}
In the practice, the transactions encoded by the inputs $u$ are processed with a strict ordering that can be enforced with only the sender's state
\begin{equation}
    u_{(i,j)} \le x_i \label{absoluteflows}
\end{equation}
as in definition \ref{transactionblock}. The model in equation \eqref{netflows} is a relaxation of the enforced requirement in equation \eqref{absoluteflows}; any block comprised of actions $u(k)$ that respect the individual transaction validity requirement equation \eqref{absoluteflows} will satisfy the conservation law in equation \eqref{netflows}. The relaxed equation in presented to demonstrate that the case of the Bitcoin network flow is in fact stronger than the conical network flow models in the controls literature. 

\subsection{Introducing Rewards}

Since the genesis block contained an empty state these requirements would make for a trivial trajectory. In order to introduce funds into the economy, a driving function $M(k)=\mu_k v(k)$ is added:
\begin{equation}
x(k+1) = A_k x(k) + B_k u(k) + \mu_k v(k).
\end{equation}
The function $M(k)\in\mathbb{R}^{n_{k+1}} $ is decomposed into a scheduled positive scalar reward $\mu_k \in \mathbb{R}_+$ and a stochastic vector $v(k)\in \mathbb{R}^{n_{k+1}}_+$ such that $\sum_i v_i(k)=1$. The vector $v(k)$ denotes the distribution of the mining rewards across all accounts including potential allocation to new accounts or may be distributed by any arbitrary rule across an arbitrary subset of accounts, such as a mining pool.

Another key property of the Bitcoin is again recovered from out state space model. Define a scalar subspace of the state
\begin{equation}
y(k) = \mathbf{1}'x(k) =\sum_i x_i(k).
\end{equation}
In order to under stand the behavior of this output function, consider
\begin{equation}
    x(k)-A_{k-1}x(k-1) = B_k u(k) + \mu_k v(k)
\end{equation}
and construct the state by summing the history of changes
\begin{equation}
\begin{split}
    x(K) = & A_{K-1}\cdots A_0 x(0)\\
    &+ \sum_{k=1}^{K-1} A_{K-1}\cdots A_{k} \left(x(k)-A_{k-1}x(k-1)\right)
\end{split}
\end{equation}
becomes
\begin{equation}
\begin{split}
    x(K) = & A_{K-1}\cdots A_0 x(0)\\ & + \sum_{k=1}^{K-1} A_{K-1}\cdots A_{k} \left(B_k u(k) + \mu_k v(k)\right).
    \end{split}
\end{equation}
When this expression is used to compute
\begin{equation}
    \begin{split}
    y(k) = &\mathbf{1}' x(K) \\
    =& \mathbf{1}'A_{K-1}\cdots A_0 x(0)\\ 
    &+ \sum_{k=1}^{K-1} \mathbf{1}'A_{K-1}\cdots A_{k} \left(B_k u(k) + \mu_k v(k)\right).
    \end{split}
\end{equation}
Since $x(0)=0$, it follows from equation \eqref{sums} that
\begin{equation}
y(K) = \sum_{k=1} \mu_k
\end{equation}
when one recalls that $A_k$ is an augmented identity matrix, that $\mathbf{1}'v(k)=1$ observes that $B_k$ is an incidence matrix: $\mathbf{1}'Bu = 0$ for all $u$. 

%\textcolor{red}{... this will actually require a bit more effort to show, the expanded state space makes the telescoping sum a little harder to follow: $\Delta x(k) = (x_k-Ax_{k-1})$ is what makes it telescope properly. I want some guidance how best to expand this arguement formally.}

In the case of Bitcoin the mining rewards are on a convergent schedule ensuring the maximum total supply
\begin{equation}
y_\infty = \lim_{k\rightarrow \infty} y(k) = \sum_{k=1}^{\infty} \mu_k
\end{equation}
converges to the desired quantity. 

In the Bitcoin network the mining rewards are defined over $i=(1,\ldots,32)$ halving intervals $r_i=(k_0,k_1,\ldots,k_{209999})$ each including $210000$ blocks resulting in
\begin{equation}
    \mu_k = \frac{\lfloor{\frac{50 \cdot 10^8}{2^i}}\rfloor}{10^8} \hbox{where $k\in r_i$.}
\end{equation}
 After the $32^{nd}$ interval the minted block rewards cease and the total quantity of Bitcoin is conserved. By computing the sum over the intervals, the final sum of Bitcoin
$y_\infty = 20999999.9769$, generally quoted as $21$ million BTC. This does not account for the potential loss of control of accounts with Bitcoin balances which reduces the effective supply.

\subsection{Globally Invariant Properties from Local Rules}
%
%\textcolor{red}{... this is important and non-obvious to most people I have interacted with}
%
While the Bitcoin economic network is a somewhat trivial system to study from a dynamical systems perspective, it is actually much like biological coordination models, an example of complex global behaviors emerging from simple local rules. The critical elements of the above example are: 
\begin{itemize}
\item The trajectory of the system is defined entirely in terms of its state transitions and the initial conditions.
\item The dynamical system model remains structurally invariant even as the number of account grows unbounded.
\item The inputs or actions are completely under the control of local agents called accounts.
\item The set of legal actions for each agent are defined and verifiable with information local to those agents states.
\item The definitions of the local legal actions provide properties that are consistent with suitability as a financial ledger of record.
\item The driving function combined with the legal actions guarantee that a low dimensional global property is enforced throughout the entire trajectory.
\end{itemize}

The most powerful part about this characterization is that the system literally tracks a desired property $y(k) = \sum_k \mu_k$ for the entire trajectory, in fact in any valid trajectory, with no assumptions about the actions of the individual agents. This indicates that it is proper to think of blockchain-enabled economic systems as engineered economies where it is possible to encode the legal state transitions in such a manner as to mathematically ensure the emergence of a low dimensional global properties.

\section{Value functions, Invariants, Optimization and Control} \label{sec:Control}

Where Bitcoin was precisely characterized by the set of hard coded rules, more general Turing complete networks such as the Ethereum network are capable of much richer state spaces and legal state transitions. In this section, the concepts explored in the Bitcoin section will be generalized for use in characterizing a more general system state.

Consider that each account may contribute a set of state variables to each other account; the cardinality of the state space grows super-linearly. At first pass this may seem to make the system difficult to work with, the opposite may in fact be true. The creator of an account is likely to have goals specific to the states it instantiates and can define precisely the functions which mutate those states including relations to other states. In particular, the goals for the system are likely to exist in a much lower dimension than the state itself and thus it is totally reasonable to provide an extremely high degree of freedom to the agents in the system while formally enforcing those goals.

For the purpose of this characterization, consider a smart contract account $\alpha$ in Turing complete computational network.  
\begin{definition}
The account $\alpha$ defines a state variable $z_a$ for all accounts $a\in \mathcal{A}_k$, for which the global state 
\[
z(k) = \{z_a\in \mathcal{Z}_a | \forall a\in \mathcal{A}_k \} \in \mathcal{Z} \subset \mathcal{X},
\]
where $\mathcal{Z}_a$ is an arbitrary state space contributed to every account $a$ by account $\alpha$.  
\end{definition}

Be aware that $z(k)$ includes part of the state of every agent $x_a(k)$, not the local states of agent $\alpha$ which is denoted $x_\alpha(k)$. Furthermore, the account $\alpha$ provides a set of methods that allow any account to modify the global state $z(k)$.
\begin{definition}
The set of functions provided by account $\alpha$ is 
\begin{equation}
\mathcal{F}_{\alpha} = \{ f_l:(\mathcal{U}_l,\mathcal{Z})\rightarrow  \mathcal{Z})  | \forall l\}
\end{equation}
where $\mathcal{U}_l$ is the set of actions or inputs to each function $f_l$ and $\mathcal{Z} \subset \mathcal{X}$ is the domain. \end{definition}

These functions define the feasible trajectories in the state of the system. Engineering these systems involves formally determining these functions in order to ensure system level properties are met. System level properties can be characterized by scalar functions of the state. The system becomes increasingly more complex as contracts are introduced to build on each others state variables. 

However, building on previous contracts is accomplished by calling methods previously available made available to mutate their states, it suffices for now to examine properties in the context of a single contract; that is to say, assume all states and state transitions are provided by account one account $\alpha$. Then $z(k)=x(k)$ be the full state with domain $\mathcal{Z} =\mathcal{X}$ and $\mathcal{F}_{\alpha}=\mathcal{F}$ accounts for all valid state transitions.

\subsection{Value Functions}

Value functions are output functions engineered to encode desired global properties. By limiting them to be positive scalars additional mathematical equipment can be brought to bear regarding convergence properties.

\begin{definition}
A Value function is a non-negative scalar function of the state of the economic network
\[
V: \mathcal{X} \rightarrow \mathbb{R}_+
\]
encoding some property.
\end{definition}

These functions are measures of properties and need to be constructed for a particular property. The first case, that will be considered is when $V(x)=c$ encodes an equality constraint. It is possible to achieve such a constraint within the network by proving an invariant.

\begin{definition} \label{invariant}
An invariant property $V(x)=c$ is ensured by verifying that
\[
V(f_l(u,x)) = V(x)
\]
for all functions $f_l\in \mathcal{F}$, actions $u\in \mathcal{U}_l$ and all legal states $x\in\mathcal{X}$.
\end{definition}
\begin{figure}[h]
\includegraphics[width=6cm]{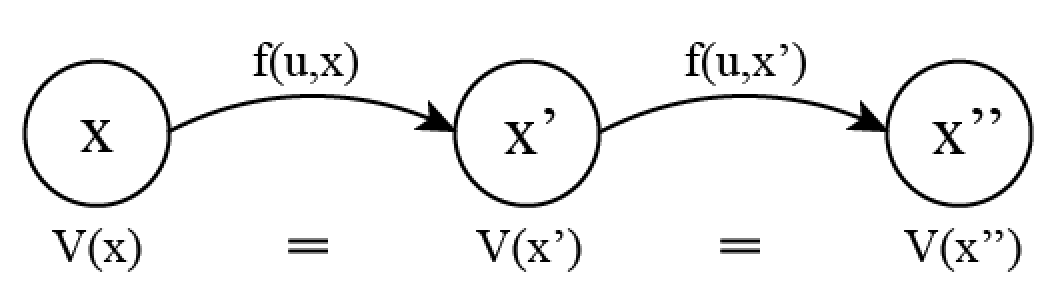}
\centering
\caption{Illustration of invariant properties. $V$ is invariant under all $f(u, x)$ for all valid actions $u$. }
\centering
\end{figure}

In the definition above, all states are considered but as needed the same definition can be used over an arbitrary subset set of the state space. In practice, the engineer will find it prudent to work with a minimal subset of the state. 

The fewer states which actually appear in the domain of $V(x)$ and $f_l(x)$ the more easily the invariant property in Definition \ref{invariant} will be to demonstrate. Further note that the construction of $V(x)$ allows a large class of set based constraints. 

In the Bitcoin example the positivity requirement could be encoded as $V(x) = -\sum_i \min\{0,x_i\}$=0. This function is well defined for all $x\in \mathbb{R}^{n}$, but is equal to zero only if every balance is positive. The fact that all send actions $u_e$ require that $x_i>u_e$ where $e=(i,j)$ guarantees this invariant. 

\subsection{Maximization}
A similar construction will allow other dynamics to be enforced. In the case of a value function $V(x)$ which the economic network wishes to collectively maximize, the following construction is sufficient.

\begin{definition}
A global maximization property over value function $V(x)$ is guaranteed by the requirement
\[
V(f_l(u,x)) \ge V(x)
\]
for all functions $f_l\in \mathcal{F}$, actions $u\in \mathcal{U}_l$ and all legal states $x\in\mathcal{X}$.
\end{definition}

Observe that this does not guarantee infinite growth of $V(x)$ but rather that $V(x)$ cannot be reduced by any legal action provided. As in the case of invariant properties, it is prudent to restrict the subset of the state which your smart contract maintains as such arguments may break down in the case where other accounts provide functions that operate on these state variables in a manner which violates the property. A strict version of this mechanism can be imposed if 
\begin{equation}
V(f_l(u,x)) \ge (1+\epsilon)V(x)
\end{equation} 
for $\epsilon>0$. This is the type of construct that can be used to ensure that perpetual revenues can be produced by the ongoing use of a network. However, if these profits are not fairly aligned with the services provided in facilitating transactions, transactions may halt and from the perspective of this system, time will have stopped and there will be no more revenue.

\subsection{Value Functions as Lyapunov Functions}

A Lyapunov function is analogous to a potential function and is a tool for proving stability around an equilibrium in a dynamical system. In the case of blockchain-enabled economic systems, the dynamics are entirely defined by the states and methods provided by accounts. As a result one need not discover Lyapunov functions to prove convergence to desired properties but can instead engineer the functions provided working backwards from a value function of choice. This approach is more powerful even than invariant simple invariant properties because the system will persistently move towards the desired state even if some unforeseen action causes a change in $V(x)$. 
\begin{definition}
An exponentially globally convergent property $V(x)=0$ is guaranteed by the requirement 
\[
V(f_l(u,x)) \le \gamma V(x)
\]
for all functions $f_l\in \mathcal{F}$, actions $u\in \mathcal{U}_l$ and all legal states $x\in\mathcal{X}$ where $\gamma\in[0,1)$ is the exponential convergence rate. 
\end{definition}
These results are simple consequences of the Brouwer fixed-point theorem \cite{smart1980fixed} provided that the space $\mathcal{X}$ is a convex compact set. Note that creating such a construction is non-trivial, but it formally ensures that all legal actions will drive the system towards the target subset of the state $\{x\in \mathcal{X} | V(x)=0\}$.  This construction further implies that crypto-economic networks are well suited to Lyapunov style control.

\begin{definition}\label{control}
Consider an output function $y = g(x)$ and target trajectory $y(k)$, define a Value function $V(x)$ such that $V(x)= 0$ if and only if $g(x) = y(k)$. This construction will be denoted a Lyapunov controller provided that 
\[
V(f_l(u,x)) \le \gamma V(x)
\]
for all functions $f_l\in \mathcal{F}$, actions $u\in \mathcal{U}_l$ and all legal states $x\in\mathcal{X}$ where $\gamma\in[0,1)$ is the exponential convergence rate. 
\end{definition}
The resulting system will under all legal actions continuously attempt to converge to the subset of the state space here $g(x) = y(k)$; note that the exponential convergent rate must be faster than the evolution of the target signal $y(k)$ in order for this process to track the desired target within a reasonable neighborhood.

Returning to the Bitcoin example, enforcing the invariant $y(K) = \sum_i x_i(K) = \sum_{k=1}^K \mu_k$ is actually an explicit case of controlling the system to track a low dimensional target state that evolves over time using a positive scalar value function. No convergence argument as in Definition \ref{control} because the value function was invariant under all legal state transitions as in Definition \ref{invariant}.

It will be interesting to see what types of emergent coordination could be engineered in economic systems by leveraging Lyapunov functions in the design of the legal action sets. It is expected that more general conservation equations may encoded to properly account for financial value changing forms, similar to how conservation of energy is handled in physical systems.

\section{Conclusion and Future Considerations} \label{sec:Conclusions}

This document builds a bridge between dynamical systems theory and blockchain-enabled economic systems. In particular, we propose a state-space representation of the economic system in terms of linear time-expanding system. This novel representation allows us to use a plethora of powerful tools developed in the context of control theory for the analysis and design of blockchain-enabled systems. In particular, we propose the use of Lyapunov-like functions, originally developed in the context of dynamical systems, to properly engineer economic systems with provable properties.  A few comments are in order: First, no direct attention is paid to the actions of individuals within the network and all guarantees proposed are derived from the spaces of actions those individuals might take. Second, it may not always be possible to sufficiently limit the action spaces to ensure desired properties without considering individual incentives. In addition to the definitions provided above the author posits that value functions can be used to analyze the alignment of incentives of individuals actions with global objectives by comparing the gradients of local incentive functions with those of the global value functions. 

%%Important but we should save this topic for a future paper -- 
%Another important distinction between incentives and reachable subspaces pertains to additive compose-ability. When a new contract adds to the state and action space of an ecosystem, the reachable states of the existing contracts may not become reachable because composition is accomplished by calling the methods provided by the original contract and by referencing the existing states. A contract which is secure in the configuration sense loses no functionality through the addition of the contract. However, behaviors which which are merely incentivized could be gamed a large scale. Returning to the our controls contract, the existing contract could be viewed as a plant of physical system and the new contract, if engineered according to closed loop controls principles could potentially modify the incentive profile to trend towards a different minimum by overlaying the original potential field with another one.

A key aspect of our future research will include the design and development of simple on-chain controllers and the empirical study of the evolution of the states in the Ethereum ecosystem for comparison with theoretical principles outlined above. Tools for streamlined access to data from blockchain economic networks are under development.

\bibliographystyle{unsrt}
\bibliography{blockchain}

\begin{thebibliography}{10}

\bibitem{nakamoto2008bitcoin}
Satoshi Nakamoto.
\newblock Bitcoin: A peer-to-peer electronic cash system.
\newblock \url{https://bitcoin.org/bitcoin.pdf}, Dec 2008.
\newblock Accessed: 2015-07-01.

\bibitem{dai1998bmoney}
Wei Dai.
\newblock bmoney.
\newblock \url{http://www.weidai.com/bmoney.txt}, 1998.
\newblock Accessed: 2016-04-31.

\bibitem{back2002hashcash}
Adam Back et~al.
\newblock Hashcash-a denial of service counter-measure.
\newblock \url{http://www.hashcash.org/papers/hashcash.pdf}, 2002.
\newblock Accessed: 2016-03-09.

\bibitem{szabo2005bitgold}
Nick Szabo.
\newblock Bit gold.
\newblock \url{http://unenumerated.blogspot.co.at/2005/12/bit-gold.html}, 2005.
\newblock Accessed: 2016-04-31.

\bibitem{bentov2014cryptocurrencies}
Iddo Bentov, Ariel Gabizon, and Alex Mizrahi.
\newblock Cryptocurrencies without proof of work.
\newblock \url{http://arxiv.org/pdf/1406.5694.pdf}, 2014.
\newblock Accessed: 2016-03-09.

\bibitem{castro1999practical}
Miguel Castro, Barbara Liskov, et~al.
\newblock Practical byzantine fault tolerance.
\newblock In {\em OSDI}, volume~99, pages 173--186, 1999.

\bibitem{ethereum-pos}
{Ethereum community}.
\newblock Proof of stake: Faq.
\newblock \url{https://github.com/ethereum/wiki/wiki/Proof-of-Stake-FAQ}.

\bibitem{bitshares-dpos}
{Bitshares community}.
\newblock Delegated proof-of-stake consensus.
\newblock
  \url{https://bitshares.org/technology/delegated-proof-of-stake-consensus/}.

\bibitem{consensus-review}
Zhang Zhe Wang Xiangwei Chen~Qijun Du~Mingxiao, Ma~Xiaofeng.
\newblock A review on consensus algorithm of blockchain.
\newblock {\em IEEE International Conference on Systems, Man, and Cybernetics},
  2017.

\bibitem{bartoletti2017analysis}
Massimo Bartoletti and Livio Pompianu.
\newblock An analysis of bitcoin op\_return metadata.
\newblock \url{https://arxiv.org/pdf/1702.01024.pdf}, 2017.
\newblock Accessed: 2017-03-09.

\bibitem{factom}
{Factom}.
\newblock A practical blockchain solution for preserving, ensuring and
  validating digital assets.
\newblock \url{https://www.factom.com/}.

\bibitem{omni-layer}
{Omni Layer}.
\newblock Omni layer, an open-sourced, fully-decentralized asset platform on
  the bitcoin blockchain.
\newblock \url{http://www.omnilayer.org/}.

\bibitem{buterin2014ethereum}
Vitalik Buterin.
\newblock Ethereum: A next-generation smart contract and decentralized
  application platform.
\newblock \url{https://github.com/ethereum/wiki/wiki/White-Paper}, 2014.
\newblock Accessed: 2016-08-22.

\bibitem{bat-token}
{Brave Software}.
\newblock Basic attention token (bat) blockchain based digital advertising.
\newblock
  \url{https://www.basicattentiontoken.org/BasicAttentionTokenWhitePaper-4.pdf},
  3 2018.

\bibitem{zrx-token}
{Will Warren, Amir Bandeali}.
\newblock 0x: An open protocol for decentralized exchange on the ethereum
  blockchain.
\newblock \url{https://icowhitepapers.info/data/0x_white_paper.pdf}.

\bibitem{gridplus-token}
{Grid+}.
\newblock Grid+: welcome to the future of energy.
\newblock \url{https://gridplus.io/assets/Gridwhitepaper.pdf}.

\bibitem{szabo1994smart}
Nick Szabo.
\newblock Smart contracts.
\newblock \url{http://szabo.best.vwh.net/smart.contracts.html}, 1994.
\newblock Accessed: 2016-08-22.

\bibitem{szabo1996smart}
Szabo Nick.
\newblock Smart contracts: Building blocks for digital markets.
\newblock \url{http://szabo.best.vwh.net/smart_contracts_2.html}, 1996.
\newblock (Accessed on 08/22/2016).

\bibitem{1524055}
D.~B. DeFigueiredo and E.~T. Barr.
\newblock Trustdavis: a non-exploitable online reputation system.
\newblock In {\em Seventh IEEE International Conference on E-Commerce
  Technology (CEC'05)}, pages 274--283, July 2005.

\bibitem{Dandekar:2011:LCN:1993574.1993597}
Pranav Dandekar, Ashish Goel, Ramesh Govindan, and Ian Post.
\newblock Liquidity in credit networks: A little trust goes a long way.
\newblock In {\em Proceedings of the 12th ACM Conference on Electronic
  Commerce}, EC '11, pages 147--156, New York, NY, USA, 2011. ACM.

\bibitem{schwartz2014ripple}
David Schwartz, Noah Youngs, and Arthur Britto.
\newblock The ripple protocol consensus algorithm.
\newblock \url{https://ripple.com/files/ripple_consensus_whitepaper.pdf}, 2014.
\newblock Accessed: 2016-08-08.

\bibitem{mazieres2015modeling}
David Mazieres.
\newblock The stellar consensus protocol: A federated model for internet-level
  consensus.
\newblock \url{https://www.stellar.org/papers/stellar-consensus-protocol.pdf},
  2015.
\newblock Accessed: 2016-08-01.

\bibitem{sontag2013mathematical}
Eduardo~D Sontag.
\newblock {\em Mathematical control theory: deterministic finite dimensional
  systems}, volume~6.
\newblock Springer Science \& Business Media, 2013.

\bibitem{khalil1996noninear}
Hassan~K Khalil.
\newblock Nonlinear systems.
\newblock {\em Prentice-Hall, New Jersey}, 2(5):5--1, 1996.

\bibitem{smart1980fixed}
David~Roger Smart.
\newblock {\em Fixed point theorems}, volume~66.
\newblock CUP Archive, 1980.

\end{thebibliography}

\end{document}